\documentclass[twoside,a4paper,11pt]{sca}
\usepackage{graphicx}
\usepackage{hyperref}
\usepackage{movie15}
\usepackage{natbib}  
\begin{document}
\pagenumbering{arabic}
\pagestyle{myheadings}
\thispagestyle{empty}
{\flushright\includegraphics[width=\textwidth,bb=90 650 520 700]{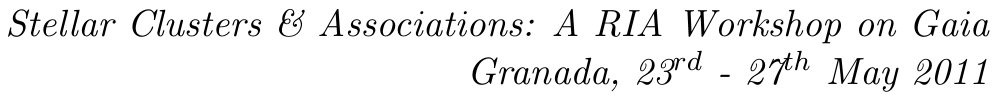}}
\vspace*{0.2cm}
\begin{flushleft}
{\bf {\LARGE
%
Do all O stars form in star clusters?
%
}\\
\vspace*{1cm}
%
Carsten Weidner$^{1}$,
Vasilii V.~Gvaramadze$^{2,3,4}$,
Pavel Kroupa$^{4}$, and
Jan Pflamm-Altenburg$^{4}$
%
}\\
\vspace*{0.5cm}
%
$^{1}$
Scottish Universities Physics Alliance (SUPA), School of Physics \& Astronomy, University of St Andrews, 
North Haugh, St Andrews, Fife KY16 9SS, Scotland, UK\\
$^{2}$Sternberg Astronomical Institute, Moscow State University, Universitetskij Pr. 13, Moscow 119992, Russia\\
$^{3}$Isaac Newton Institute of Chile, Moscow Branch, Universitetskij Pr. 13, Moscow 119992, Russia\\
$^{4}$Argelander-Institut f\"ur Astronomie (Sternwarte), Auf dem H{\"u}gel 71,
D-53121 Bonn, Germany
%
\end{flushleft}
%
\markboth{
Do all O stars form in star clusters?
}{ 
%
Weidner et al.
%
}
\thispagestyle{empty}
\vspace*{0.4cm}
\begin{minipage}[l]{0.09\textwidth}
\ 
\end{minipage}
\begin{minipage}[r]{0.9\textwidth}
\vspace{1cm}
\section*{Abstract}{\small
%
The question whether or not massive stars can form in isolation or only in star clusters is of great importance for the theory of (massive) star-formation as well as for the stellar initial mass
function of whole galaxies (IGIMF-theory). While a seemingly easy question it is rather difficult to answer. Several physical processes (e.g.\,star-loss due to stellar dynamics or gas expulsion) and
observational limitations (e.g.\,dust obscuration of young clusters, resolution) pose severe challenges to answer this question. In this contribution we will present the current arguments in favour and against the idea that all O stars form in clusters.
%
\normalsize}
\end{minipage}
%
%
%
\section{Introduction \label{intro}}
Besides huge efforts observationally and theoretically, the question how massive stars (m $>$ 10 $M_\odot$) 
form has not been answered satisfactory yet. Amongst others, two main theories of the formation of massive
stars have been put forward. One theory being the monolithic collapse of very massive and dense cores 
\citep{MT03,K06} into a single star or a binary. As this theory has no requirements for the surroundings of such 
cores, it allows the formation of isolated O stars without an star cluster associated to them. However, the 
question how the isolated massive core can form remains open. The other important
theory is competitive accretion, where massive stars form in the dense centres of star clusters \citep{BBC97} and
therefore it does not predict isolated O stars.

\begin{figure}
\center
\includegraphics[width=9cm]{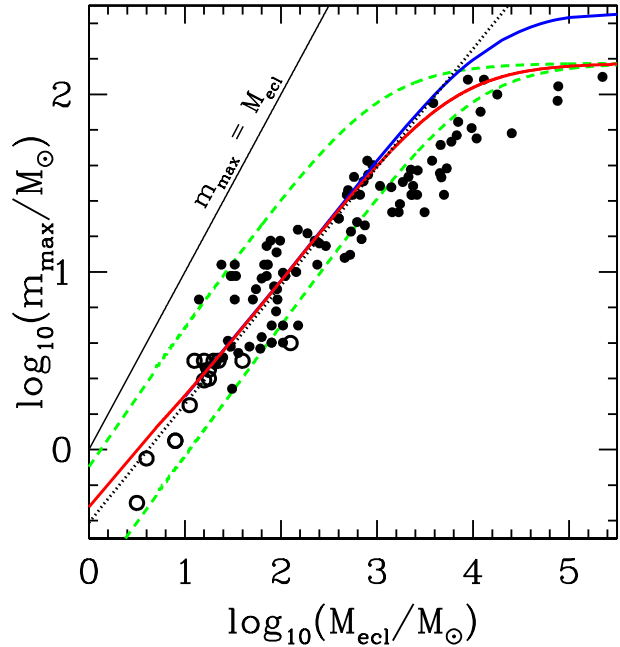} 
\vspace*{-2cm}
\caption{\label{fig1} Most-massive star ($m_\mathrm{max}$) in a cluster versus the stellar mass of the young 
dynamically unevolved "embedded" cluster ($M_\mathrm{ecl}$). The filled
dots are observations compiled by \citet{WKB09}, while the open circles are new data from \citet{KM10}. The 
solid lines through the data points are the medians expected for random sampling when using a fundamental 
upper mass limit, $m_\mathrm{max*}$, of 150 $M_\odot$ (lower solid, red line) and $m_\mathrm{max*}$ = 300 
$M_\odot$ \citep[upper solid, blue line;][]{CSH10}. The dashed lines are the 1/6 and 5/6th quantiles which 
encompass 66\% of the most-massive stars if they are randomly sampled from an IMF with an upper mass limit of
150 $M_\odot$. The dotted line shows the prediction for a relation by \citet{BBV03} from numerical models of 
molecular clouds with less than 10000 
$M_\odot$. The thin solid line marks the limit where a cluster is just made out of one star.
}
\end{figure}

Shown in Fig.~\ref{fig1} is a compilation of star clusters from the literature and the most-massive stars
in them \citep{WK05b,WKB09}. The Figure shows a tight correlation between the mass of the most-massive
star, $m_\mathrm{max}$, and the mass of the cluster, $M_\mathrm{ecl}$. Especially for massive clusters
this $m_\mathrm{max}$-$M_\mathrm{ecl}$ relation is well below what is expected when randomly sample
stars (red solid line) from an stellar initial mass function (IMF). If the formation of massive stars is independent
of their natal environment, the $m_\mathrm{max}$-$M_\mathrm{ecl}$ relation should be well described by
random sampling. The existence of a non-trivial $m_\mathrm{max}$-$M_\mathrm{ecl}$ relation is therefore
direct evidence against the formation of O stars in isolation.

But if all O stars are formed in star clusters why do we indeed observe O stars in the Galactic field?

\section{Massive field stars}
Massive field stars are OB stars that are not members of any currently known star cluster, OB association or
star-forming region. So far about 20-30\% of all Galactic O stars are in the field \citep{G87}. These stars can
be separated into two groups:
\begin{itemize}
\item $\sim$ 25\% are high-velocity OB stars \citep[typical $>$ 30 km s$^{-1}$; runaway stars;][]{B61,G87},
\item $\sim$ 75\% are low-velocity ($\le$ 30 km s$^{-1}$) OB stars.
\end{itemize}
Though, have these massive field stars formed in isolation or could they originate from star clusters?

\subsection{The origin of runaway OB stars}
Generally, high velocity massive field stars OB stars (runaway stars) originate from two mechanisms:

\begin{itemize}
\item Disruption of a short-period binary after a supernova explosion \citep{B61,S91},
\item or by three- or many-body interactions in star clusters \citep{PRA67,GB86}.
\end{itemize}

Recently, this picture has been expanded by a third mechanism which combines the previously known two
ones. In the two-step ejection mechanism, a binary of two O stars is ejected and than later the more massive
star in the binary explodes as a supernova, shooting the secondary in a random direction and changing its
velocity, too \citep{PK10}. Such stars can not be traced back to their parent star cluster and will be mistakenly
identified as high-mass stars formed in isolation.

\begin{figure}
\center
\includegraphics[width=9cm]{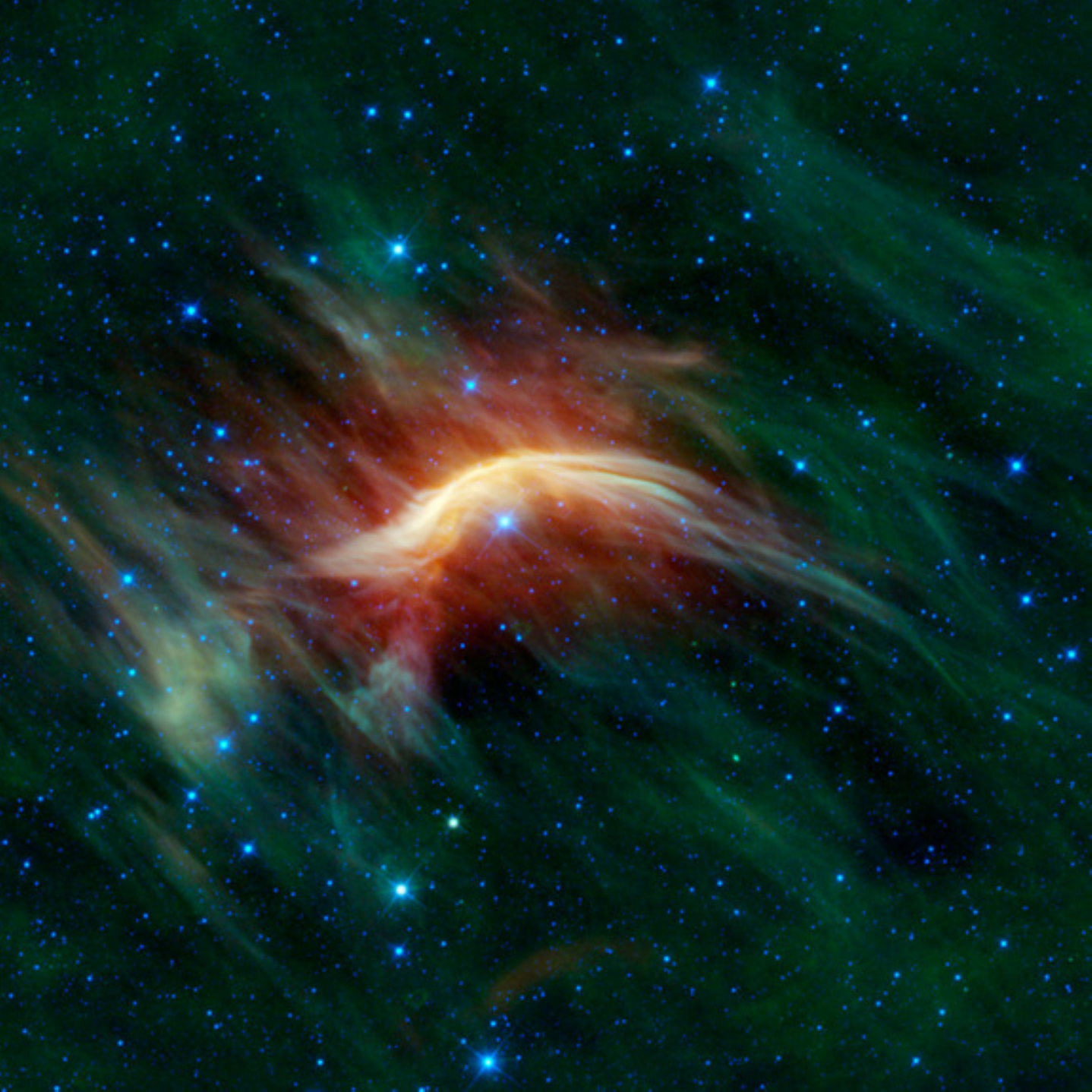} 
\caption{\label{fig2} Bow-shock in front of $\zeta$ Ophiuchi. Credit: NASA/JPL-Caltech/WISE team.
}
\end{figure}

Runaway stars can be observationally identified by several direct and indirect methods. The direct methods 
are based on detection of high ($>$ 30 kms$^{-1}$) peculiar transverse and/or radial velocities via proper 
motion measurements \citep[e.g.][]{B61,MMS98,MC05,TNH10} and spectroscopy 
\citep[e.g.][]{MPP05,ELS06,EWC10}, respectively. The indirect 
indications of the runaway nature of some field OB stars are the large ($>$ 200 pc) separation of these stars from
the Galactic plane \citep[e.g.][]{B61,Bl93} and the presence of bow shocks around them 
\citep[e.g.][]{GB08,GKP10,GPK10}. 
Fig.~\ref{fig2} shows an example of such a bow shock.

\subsection{The origin of low-velocity field OB stars}
For low-velocity massive field stars several different origins are possible:
\begin{itemize}
\item They formed in isolation,
\item they are the low-velocity tail of the ejected stars,
\item there true 3D-velocity has not been determined correctly and they are unrecognised runaways,
\item they are members of undetected or dissolved star clusters, 
\item they are blue stragglers, either directly ejected as such or formed from merged ejected binaries.
\end{itemize}

It is important to note that a two-step ejection can slow-down or stop runaway OB stars \citep{PK10} and
therefore let them look like low-velocity field OB stars.

\begin{figure}
\center
\includegraphics[width=8cm]{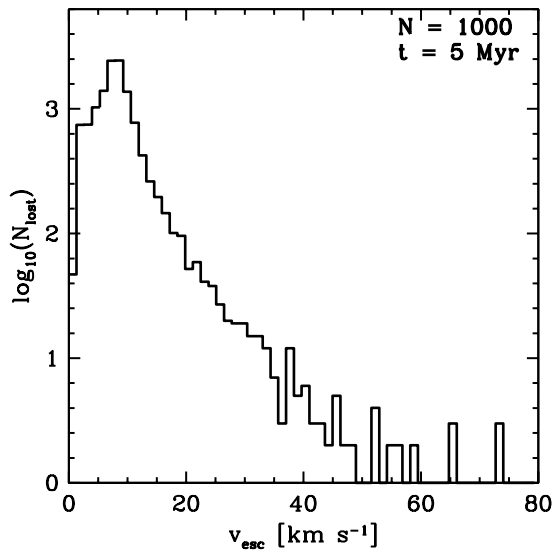} 
\vspace*{-2.2cm}
\caption{\label{fig3} Velocity distribution of all stars (including low-mass stars) escaped after 5 Myrs from 100 star 
cluster calculations starting with 500 binaries ($M_\mathrm{cluster}$ $\sim$ 340 $M_\odot$) each 
\citep{WBM10}. Note that the velocity distribution and the escape fraction depends strongly on the initial stellar 
density of the cluster (see the contribution by Pflamm-Altenburg this volume).
}
\end{figure}

Furthermore, is the definition of runaway stars misleading. The typical velocity of 30 km s$^{-1}$ used to 
separate runaways from low-velocity stars is an observational definition and not necessarily a typical 
ejection velocity of stars from a star cluster. Shown in Fig.~\ref{fig3} is the distribution of escape velocities
of all ejected stars form a series of 100 $N$body calculations of 500 binaries over 5 Myr. A high-velocity
tail certainly exists but in such low-mass clusters these are not massive stars. In total over all 100 $N$body
calculations 30 stars ($\sim$10\% of the massive stars, each individual cluster has only 3 stars above 10 
$M_\odot$) with masses above 10 $M_\odot$ have been ejected 
within 5 Myr. But their escape velocities are only between 7 and 14 km s$^{-1}$. Though, they still travel 
between 1 and 50 pc after their escape within the 5 Myr \citep{WBM10}.

\section{High-mass star-formation in isolation?}
From an originally proposed fraction of 4\% of O stars formed in isolation \citep{DTP04,DTP05} careful 
back-tracking of individual stars \citep{SR08} could reduce this percentage to 2\%. Further studies of the
five remaining candidates revealed that two of them have bow shocks and are therefore ejected stars, too 
\citep{GB08}. This reduces the fraction of O stars formed in isolation to 1\% (Weidner et al.~2011b, submitted).
 But it should be noted that of the three remaining
candidates two cases (HD 193793 and HD 202124) haven't been studied for bow shocks so far. And while in 
final the case of HD 124314 no bow shock has been found it is important to note that bow shocks are only 
formed under certain circumstances. Furthermore, does the two-step ejection method predict between 1 and
2\% of apparently isolated O stars which can not be tracked back to their natal cluster \citep{PK10}.

To conclude, there is no significant evidence for massive stars formed in isolation.

%
%
\small  
%
\section*{Acknowledgments}   
%
VVG acknowledges financial support from the Deutsche Forschungsgemeinschaft.

%
%
%
%
%

\bibliographystyle{aa}
\bibliography{mnemonic,Weidner_C}

\end{document}